\documentclass[aps,english,superscriptaddress,preprintnumbers,reprint,footinbib,amsmath,amssymb,pra]{revtex4-2}
\usepackage{graphicx}
\usepackage[version=3]{mhchem}
\usepackage{graphicx}
\usepackage{dcolumn}
\usepackage{bm}
\usepackage{hyperref}
\hypersetup{
    colorlinks=true,
    linkcolor=blue,
    filecolor=blue,
    urlcolor=blue,
   citecolor=blue,
}
\usepackage{enumitem}
\usepackage{siunitx}

\begin{document}

\title{Reshaping the Jaynes-Cummings ladder with Majorana bound states}

\author{L. S. Ricco}
\email[corresponding author: ]{lsricco@hi.is}
\affiliation{Science Institute, University of Iceland, Dunhagi-3, IS-107,
Reykjavik, Iceland}

\author{V. K. Kozin}
\affiliation{Department of Physics, University of Basel, Klingelbergstrasse 82, CH-4056 Basel, Switzerland}
\affiliation{Science Institute, University of Iceland, Dunhagi-3, IS-107, Reykjavik, Iceland}

\author{A. C. Seridonio}
\affiliation{S\~ao Paulo State University (Unesp), School of Engineering, Department of Physics and Chemistry, 15385-000, Ilha Solteira-SP, Brazil}

\author{I. A. Shelykh}
\affiliation{Science Institute, University of Iceland, Dunhagi-3, IS-107,
Reykjavik, Iceland}
\affiliation{Department of Physics, ITMO University, St.~Petersburg 197101, Russia}

\date{\today}

\begin{abstract} 
We study the optical properties of a hybrid device composed by a quantum dot (QD) resonantly coupled to a photonic mode of an optical microcavity and a Majorana nanowire:  a topological superconducting segment hosting Majorana bound states (MBSs) at the opposite ends. In the regime of strong light-matter coupling, it is demonstrated that the leakage of the Majorana mode into the QD opens new optical transitions between polaritonic states formed due to hybridisation of material excitation with cavity photons, which leads to the reshaping of the Jaynes-Cummings ladder and can lead to the formation of a robust single-peak at cavity eigenfrequency in the emission spectrum. Moreover, weak satellite peaks in the low and high frequency regions are revealed for the distinct cases of highly isolated MBSs, overlapped MBSs and MBSs not well localized at the nanowire ends.  
\end{abstract}


\maketitle

\section{Introduction}
Almost a century ago, Ettore Majorana discovered the representation of the Dirac equation having real wave functions as its solutions~\cite{Majorana1937}. They describe exotic particles which are equivalent to their own antiparticles, known as Majorana fermions. In the last years, the Majorana's proposal  attracted continuous attention of the researchers outside high energy physics community, once it was shown that Majorana quasiparticles can emerge in condensed matter systems~\cite{RevMajoranaAlicea,RevMajoranaFranz,RevMajoranaAguado,ReviewYazdani2021,JelenaReviewJAP2021}. These Majorana-like excitations arise in topologically protected phases of matter~\cite{RevMajoranaAguado} and possess exotic non-Abelian statistics~\cite{RevNonabelian2008}. Besides the interest from the fundamental viewpoint, the remarkable features of the Majorana quasiparticles make them attractive potential candidates for performing decoherence-free quantum computing operations~\cite{Kitaev2003,AasemPhysRevX.6.031016(2016)}.   

Among several platforms where Majorana quasiparticles can emerge~\cite{RevMajoranaAlicea,ReviewYazdani2021}, one-dimensional hybrid semiconducting nanowires with strong Rashba spin-orbit interaction placed in contact with a superconductor~\cite{Kitaev2001,LutchynPRL2010,OregPRL2010,LutchynReviewMat2018} has been considered as one of the best options. Under application of an external magnetic field, these so-called Majorana nanowires undergo the topological phase transition, characterized by the emergence of a \textit{p}-wave superconducting gap supporting zero-energy Majorana bound states (MBSs)  at the opposite ends of the proximitized nanowire~\cite{Kitaev2001}. 

\begin{figure}[t]
	\centerline{\includegraphics[width=3.5in,keepaspectratio]{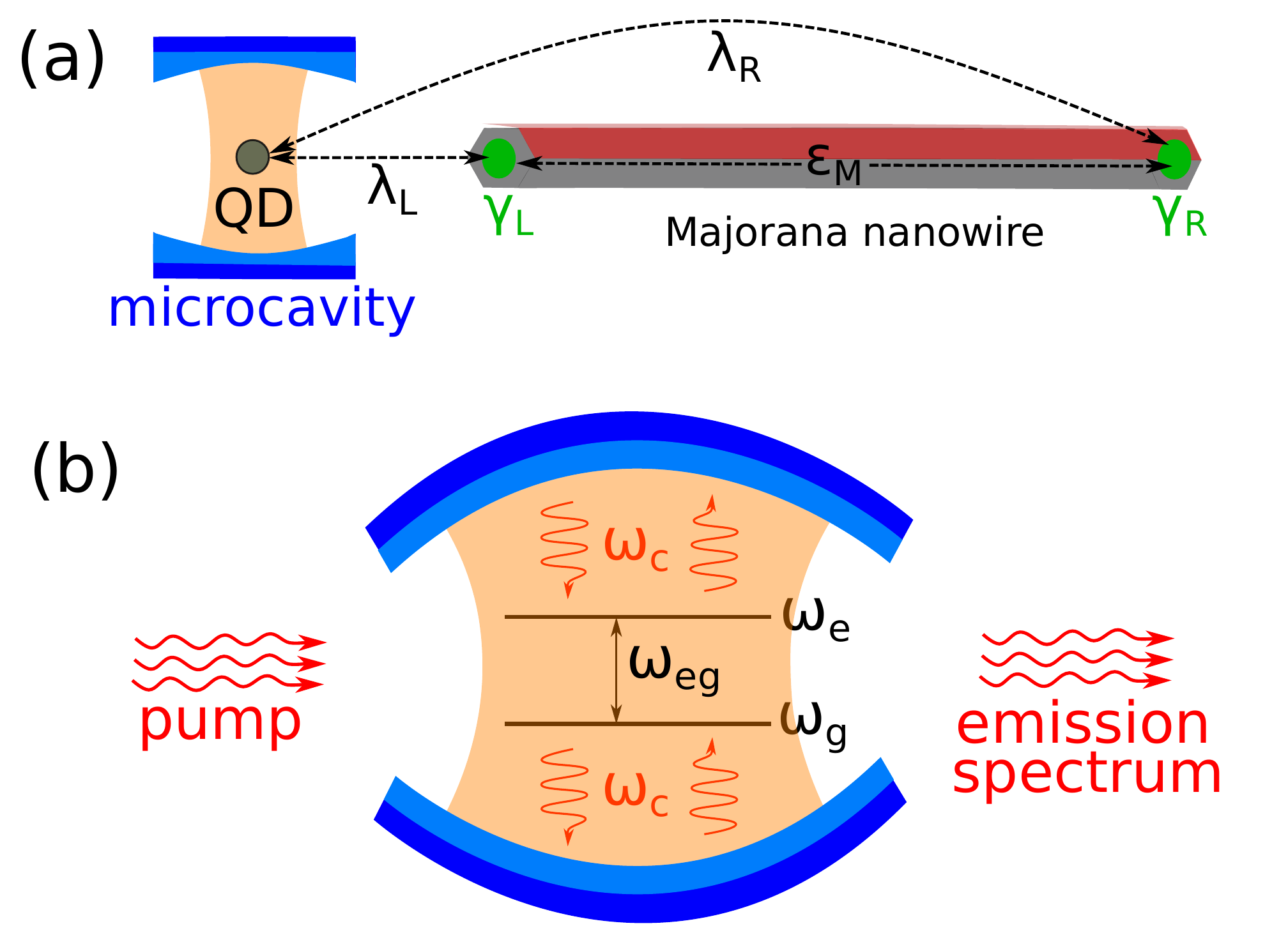}}
	\caption{\label{fig:System}(a) The sketch of the proposed device: a quantum dot (QD) embedded in an optical microcavity and coupled to a Majorana nanowire hosting Majorana bound states (MBSs) at its opposite ends (light green circles). $\lambda_{L}$ and $\lambda_{R}$ represent the couplings between the QD excited level and the left ($\gamma_{L}$) and right MBSs ($\gamma_{R}$), respectively, while $\varepsilon_{M}$ stands for the overlap amplitude between the MBSs. (b) Optical transitions for a QD placed inside an optical microcavity. The dot interacts resonantly with a single-mode photonic field of the pumped cavity with frequency $\omega_{c}=\omega_{eg}=\omega_{e} - \omega_{g}$ brought in resonance with the energy of the optical transition,  where $\omega_{e}$ and $\omega_{g}$ stand for the energies of the excited and ground states of the QD, respectively.}
\end{figure}

One of the signatures of the formation of topologically protected MBSs at the ends of a Majorana nanowire is the emergence of a quantized zero-bias conductance peak (ZBCP) in tunneling conductance through a nanowire at very low temperatures~\cite{NichelePRL2017,MourikScience2012,PradaPhysRevB.86.180503(2013),AlbrechtNature2009,KrogstrupNatMater2015,DengScience2016,Zhang2019}, which is robust to variations of the system parameters, such as gate-voltages and magnetic fields. In the corresponding geometry, a quantum dot (QD) side-coupled to a Majorana nanowire has been employed as a tunneling spectroscopic tool to probe MBSs and also measure the so-called degree of Majorana nonlocality~\cite{VernekPRB2014,DengScience2016,DengPRB2018,Prada2017,DengPRB2018,Penaranda2018,DJClarcke2017,RiccoPRB2020,SciRepIsoconductance2021}.

While tunneling spectroscopic properties of hybrid QD-Majorana nanowires have been extensively studied over the last years, their optical properties did not attract any substantial attention, although some works in this field have started to appear ~\cite{CottetPRB2013,MirceaPRL2012,DartiailhPRL2017,TangAIP2015,WangSciRep2015,ChenSciRep2018,ChiFP2020,ZantenNaturePhys2020,peng2020floquet,Ricco2021accessing,Contamin2021condmat,bittermann2021probing}. On the other hand, individual QDs are widely applied in optical quantum information processing, where they can be used as solid state source of single photons \cite{Senellart2017}, entangled photon pairs on demand \cite{Benson2000,Johne2008,Peiris2017}, and other types of non-classical light \cite{Carsten2015,Trivedi2020,Bin2020}, as well as quantum filters \cite{Ridolfo2010,Fischer2017,Foster2019} and other building blocks of quantum integrated photonic circuits \cite{Arcari2014,Lodahl2015,Dietrich2016}.

In this context, a system consisting of a QD embedded inside a photonic cavity is of particular interest. If the cavity have a high quality factor and the excitonic transition in the dot is brought in resonance with the photonic eigenfrequency, the system enables a dramatic enhancement of the light-matter interaction. In this situation, the regime of strong light-matter coupling is achieved \cite{Yoshie2004,Reithmaier2004,Englund2007}, and light and matter quantum states become hybridized.  
The energy spectrum is dramatically reshaped, and the so called Jaynes-Cummings
ladder emerges \cite{Jaynes1963,Bruce1993,Laussy2012}. Its formation changes the emission spectrum, which for a symmetric QD evolves from the Rabi doublet to Mollow triplet \cite{Mollow1969} with increase of the pump intensity, both for resonant and non-resonant pumping schemes \cite{DelValle2011,Fischer2016,Carreno2016}. Incorporation
of the spatial inversion asymmetry into the quantum system further changes the emission pattern and leads to the opening of optical
transitions at the Rabi frequency which were forbidden in the symmetric case \cite{Kibis2009,Savenko2012}. As well, transition from the case of small dots, where material excitations are fermion-like, to large quantum dots, when excitons behave more as bosons is also associated with restructuring of the emission spectrum, which starts to reveal a complex multiplet structure \cite{Laussy2006,DelValle2008}. 

In the present work, we explore the effects of the coupling between a Majorana nanowire and a QD embedded in a microcavity (Fig.~\ref{fig:System}) in the regime of strong light-matter coupling, with the excitonic transition in the dot brought in resonance with the cavity eigenfrequency. It is demonstrated that the leakage of the MBS into the QD opens new optical transitions between the polaritonic states of the QD-microcavity system, which are originally forbidden in absence of the Majorana nanowire. More specifically, the opening of these transitions are allowed due to the reshaping of the primary Jaynes-Cummings ladder due to the QD-Majorana nanowire finite coupling. The spatial position of the MBSs with respect to the nanowire ends, as well as the overlap between each other directly affect the way in which the ladder rungs are reshaped. The emission spectrum is strongly modified due to the coupling between the QD and Majorana nanowire, showing a prominent single-peak at the cavity eigenfrequency. 

The paper is organized as follows. In Sec.~\ref{Hamiltonian}, the model Hamiltonian is introduced. In Sec.~\ref{Optical transitions}, the theory describing the optical transitions and emission spectrum is presented. In Sec.~\ref{Results and Discussion}, we show the numerical results for the new optical transitions and corresponding emission spectrum for the cases of highly isolated and overlapped MBSs. The main results are summarized in Sec.~\ref{Conclusions}.

\section{Model and Methods}\label{Model}

\subsection{The Hamiltonian}\label{Hamiltonian}

We consider the system schematically shown in Fig.~\ref{fig:System}(a). A two-level QD, with the ground state $|g\rangle$ and the excited state $|e\rangle$, is embedded inside a single-mode cavity with frequency $\omega_{c}$. In the absence of interaction, the ground state of the QD corresponds to the case with one electron in the lowest level and no electrons in the excited level of the QD, i.e, $|g\rangle\equiv|1,0\rangle$, while the excited state is the opposite: $|e\rangle\equiv|0,1\rangle$. The QD interacts resonantly with the cavity photonic field tuned at the frequency $\omega_{c}=\omega_{eg}=\omega_{e}-\omega_{g}$, where $\omega_{e}$ and $\omega_{g}$ are the energies of the excited and ground states of the dot, respectively [see Fig.~\ref{fig:System}(b)]. The QD excited level is coupled to both the MBSs at opposite ends of a superconducting nanowire in the topological phase. The full Hamiltonian which describes the system reads:
\begin{equation}
\hat{\mathcal{H}}=\hat{\mathcal{H}}_{JC}+\hat{\mathcal{H}}_{M},\label{eq:Hfull}
\end{equation}
where
\begin{eqnarray}
\hat{\mathcal{H}}_{JC}&=&\omega_{c}a^{\dagger}a+\omega_{e}d_{e}^{\dagger}d_{e} + \omega_{g}d_{g}^{\dagger}d_{g}\nonumber\\ &+&\Omega_{R}(d_{g}^{\dagger}d_{e}a^{\dagger} +d_{e}^{\dagger}d_{g}a )\label{eq:HJC}
\end{eqnarray}
is the well-known Jaynes-Cummings (JC) model, describing optical transitions in the dot within the rotating wave approximation (RWA)~\cite{Jaynes1963,Bruce1993} ( $\hbar=1$). The bosonic operator $a^{\dagger}$ ($a$) creates (annihilates) the cavity photons and $d_{j}^{\dagger}$ ($d_{j}$) creates (annihilates) an electron in the QD $j-$level, where $j=e,g$. The last term in the right-hand-side of Eq.~(\ref{eq:HJC}) describes the coupling between the cavity photons and an electron in the QD, with the coupling strength $\Omega_{R}$ (Rabi frequency). It depends on the dipole matrix element of the optical transition and the cavity geometry.

The Hamiltonian which describes the topological superconducting nanowire (Majorana nanowire) side-coupled to the excited QD state reads~\cite{ VernekPRB2014,RiccoPRB2020,Prada2017,Baranger2011,Ricco2021accessing}:
\begin{eqnarray}
\hat{\mathcal{H}}_{M}&=&\imath\varepsilon_{M}\gamma_{L}\gamma_{R}+\lambda_{L}(d_{e}-d_{e}^{\dagger})\gamma_{L}\nonumber\\
& + & \lambda_{R}(d_{e}+d_{e}^{\dagger})\gamma_{R} 
\label{eq:HMajorana}
\end{eqnarray}
where the operators $\gamma_{L}$ and $\gamma_{R}$ represent the MBSs at the left and right ends of the Majorana nanowire, respectively, with $\gamma_{a}=\gamma_{a}^{\dagger}$~\cite{RevMajoranaAguado}, obeying the anticommutation relation $\{ \gamma_{a},\gamma_{b}\}=2\delta_{a,b}$ ($a,b=L,R$). The overlap strength between the MBSs is given by $\varepsilon_{M}$, and $\lambda_{L}$ and $\lambda_{R}$ are the couplings between the electron at the QD excited level and left and right MBSs, respectively.

The effective Hamiltonian of Eq.~(\ref{eq:HMajorana}) has been widely used in previous works~\cite{Baranger2011,VernekPRB2014,Prada2017,DengPRB2018,Ricco2021accessing}, and allows us to explore the following situations:
\begin{enumerate}[label=(\roman*)]
    \item Highly isolated MBSs: this case corresponds to disorder-free and longer nanowires~\cite{pan2021quantized,pan2020disorder}, where the topologically protected MBSs are well localized at the opposite ends of the nanowire ($\lambda_R = 0$) and does not overlap with each other ($\varepsilon_M = 0$)~\cite{prada2019andreev}. 
    \item MBSs localized at the nanowire ends, but with overlap between them: as in the former case, the the wavefunctions which describe the left and right MBSs are centered at the corresponding nanowire ends ($\lambda_R = 0$). However, due to a shorter nanowire length these wavefunctions strongly overlap with each other ($\varepsilon_M \neq 0$)~\cite{prada2019andreev,AlbrechtNature2009}. 
    \item right MBS shifted from its nanowire end: distinct from previous situations, now the wavefunction of the right-MBS is not centered at its corresponding nanowire end, leading to a coupling with the QD ($\lambda_R\neq 0$, $\lambda_R\gg \varepsilon_M$)~\cite{Prada2017,DengPRB2018}. This case qualitatively emulates MBSs with partial spatial separation between them, which can appear due to inhomogeneous potentials in the nanowire~\cite{Penaranda2018,Vuik2019}. Moreover, the finite value of $\lambda_R$ is associated with the degree of Majorana nonlocality~\cite{Prada2017,Ricco2021accessing}, also known as topological quality factor~\cite{DJClarcke2017,RiccoPRB2020}. 
\end{enumerate}

It is worth emphasizing that the changing of the QD-MBSs couplings $\lambda_{L}$ and $\lambda_{R}$, as well as the MBS-MBS overlap $\varepsilon_{M}$, mimics realistic physical situations, as described above. Moreover, such quantities depend on external tunable parameters, as the Majorana nanowire chemical potential and the magnetic field strength applied perpendicularly to the nanowire~\cite{Prada2017,prada2019andreev,RiccoOscillations2018}, for instance.

It is convenient to write $\hat{\mathcal{H}}_{M}$ [Eq.~(\ref{eq:HMajorana})] in terms of canonical fermion operators $c_{M}$, which obeys usual anticommutation relations for fermions $\{ c_{M},c_{M}^{\dagger}\}=1$ and $\{ c_{M},c_{M}\}=\{ c_{M}^{\dagger},c_{M}^{\dagger}\}=0$. According to this representation, the Majorana operators are rewritten as~\cite{Ricco2019,VernekPRB2014,Baranger2011}, $\gamma_{L}=(c_{M}^{\dagger}+c_{M})/\sqrt{2}$ and $\gamma_{R}=\imath(c_{M}^{\dagger}-c_{M})/\sqrt{2}$, and thus $\hat{\mathcal{H}}_{M}$ becomes:
\begin{eqnarray}
\hat{\mathcal{H}}_{M} &=&\varepsilon_{M}n_{M}+\Lambda^{-}(d_{e}c_{M}^{\dagger}+c_{M}d_{e}^{\dagger})\nonumber\\&+&\Lambda^{+}(d_{e}c_{M}+c_{M}^{\dagger}d_{e}^{\dagger}),
\end{eqnarray}
with $n_{M}=c_{M}^{\dagger}c_{M}$ being the number operator associated to the fermionic operator given by the combination of the Majorana operators $\gamma_{L,R}$ and $\Lambda^{\pm}=(\lambda_{L}\pm\lambda_{R})/\sqrt{2}$. Although $c_{M}$ stands for a canonical operator, it has a nonlocal character, once it comes from the combination of two MBSs which are spatially far apart from each other.

It should be noted that, distinct from the case of the JC Hamiltonian~\cite{Bruce1993,Savenko2012}, $[\hat{\mathcal{H}},\hat{N}]\neq0$, i.e, the system Hamiltonian of Eq.~(\ref{eq:Hfull}) does not commute with the excitation number operator 
\begin{equation}
\hat{N}= a^{\dagger}a + d_{e}^{\dagger}d_{e} + c_{M}^{\dagger}c_{M}, \label{eq:NumberOperator}     
\end{equation}
due to the presence of the term $\Lambda^{+}(d_{e}c_{M}+c_{M}^{\dagger}d_{e}^{\dagger})$ in $\hat{\mathcal{H}}_{M}$ [Eq.~(\ref{eq:HMajorana})], indicating that the number of electron-photon excitations in the system is not conserved. This is a consequence of the superconducting nature of the Majorana nanowire, which does not conserve the number of excitations.

\subsection{Optical transitions and Emission spectrum}\label{Optical transitions}

The interaction term of the JC Hamiltonian [last one in the right-hand-side of Eq.~(\ref{eq:HJC})] corresponds to the process of transferring the electron from the ground state $|g\rangle$ to the excited state $|e\rangle$ or vice versa by absorbing or emitting one photon from or to the cavity, i.e, $|g,n\rangle \leftrightarrow |e,n-1\rangle$, with $|g(e),n\rangle=|g(e),n\rangle\otimes|n\rangle$. These processes result in the formation of hybridized states, describing the QD dressed by the cavity photons, known as upper (U) and lower (L) polaritons. If the cavity is tuned exactly in resonance with ground-to-excited state  transitions, their wavefunctions are $|U,n\rangle=(|g,n\rangle+|e,n-1\rangle)/\sqrt{2}$, $|L,n\rangle=(|g,n\rangle-|e,n-1\rangle)/\sqrt{2}$. In the regime of strong light-matter coupling, the allowed optical transitions between the upper and lower polaritons due to either emission or absorption of a cavity photon gives rise to the so-called JC ladder~\cite{Bruce1993,MicrocavitiesBook}.  

The optical transitions between the eigenstates of Eq.~(\ref{eq:Hfull}) corresponding to different occupation numbers $n$ of cavity photons can be computed by considering an exchange between the cavity photons of Fig.~\ref{fig:System} and the "outside world" (the reservoir of modes outside the system). This exchange is possible due to both: an external incoherent pump applied to the QD-microcavity system and due to photons being able to escape the microcavity due to its finite lifetime. The coupling between the closed system and the reservoir can be accounted by the following Hamiltonian 
\begin{equation}
\hat{\mathcal{H}}_{ex}=\Gamma a b^{\dagger} + \text{h.c.},\label{eq:Hex} 
\end{equation}
where $b^{\dagger}$ ($b$) creates (annihilates) a photon in the external reservoir and $\Gamma$ is the constant system-reservoir coupling strength~\cite{Savenko2012}. The transition probabilities corresponding to the emission of a photon from the cavity to an empty external reservoir are proportional to
\begin{equation}
I_{if}\sim |\langle\psi_{f,n},1_{out}|\hat{\mathcal{H}}_{ex}|\psi_{i,n},0_{out}\rangle|^{2}.\label{eq:FermiGoldenruleProbability}    
\end{equation}
where $|\psi_{i,n}\rangle$ and $|\psi_{f,n}\rangle$ are initial and final eigenstates associated to the system Hamiltonian [Eq.~(\ref{eq:Hfull})], and $0_{out}$ and $1_{out}$ represent the zero and one photon states of the reservoir. By substituting Eq.~(\ref{eq:Hex}) into Eq.~(\ref{eq:FermiGoldenruleProbability}), we obtain
\begin{equation}
I_{if}  \sim  |\langle\psi_{f,n}|a|\psi_{i,n}\rangle|^{2}. \label{eq:transitions}   
\end{equation}
Although the analysis of the optical transitions through Eq.~(\ref{eq:transitions}) is useful for discussing the opening of new transitions induced in the QD-microcavity system due to its coupling with the Majorana nanowire, it does not allow us to obtain the shape of the corresponding emission spectrum for a given incoherent pump and cavity photons lifetime.

\begin{figure*}[t]
	\centerline{\includegraphics[width=6.4in,keepaspectratio]{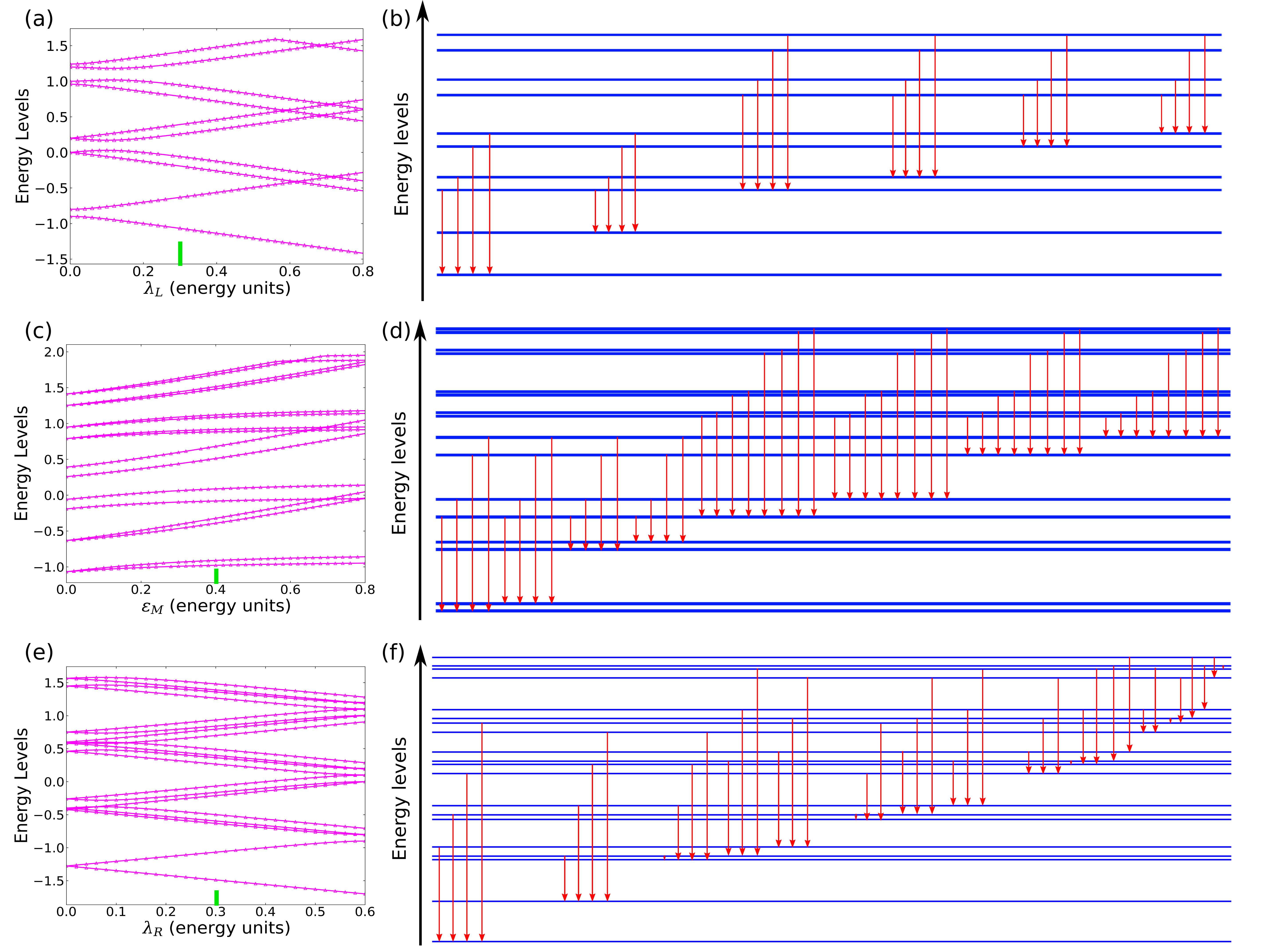}}
	\caption{\label{fig:Transitions}Left panels: Evolution of the first $20$ energy levels of the system as a function of the parameters of QD-MBSs couplings $\lambda_{L}$ and $\lambda_{R}$, as well as the overlap between the MBSs $\varepsilon_{M}$. Right panels: optical transitions between the system eigenenergies for each corresponding case. The blue lines represents the initial and final states of each corresponding transition, which is represented by a red arrow. The specific parameters adopted in each of the right panels are indicated by the light green rectangle in the corresponding left ones. (a) and (b) correspond to the case of well isolated MBSs localized at the nanowire ends, with $\lambda_{L}=0.3$ in (b). Panels (c) and (d) describe the situation of overlapped MBSs well localized at the nanowire ends, for $\lambda_{L}=0.3$ and $\varepsilon_{M}=0.4$ in (d). Panels (e) and (f) depict the case wherein the MBSs barely overlaps, but are not well localized at the ends of the nanowire, with $\lambda_{L}=0.6$ and $\lambda_{R}=0.3$ in (f).}
\end{figure*}

\begin{figure*}[t]
	\centerline{\includegraphics[width=7.2in,keepaspectratio]{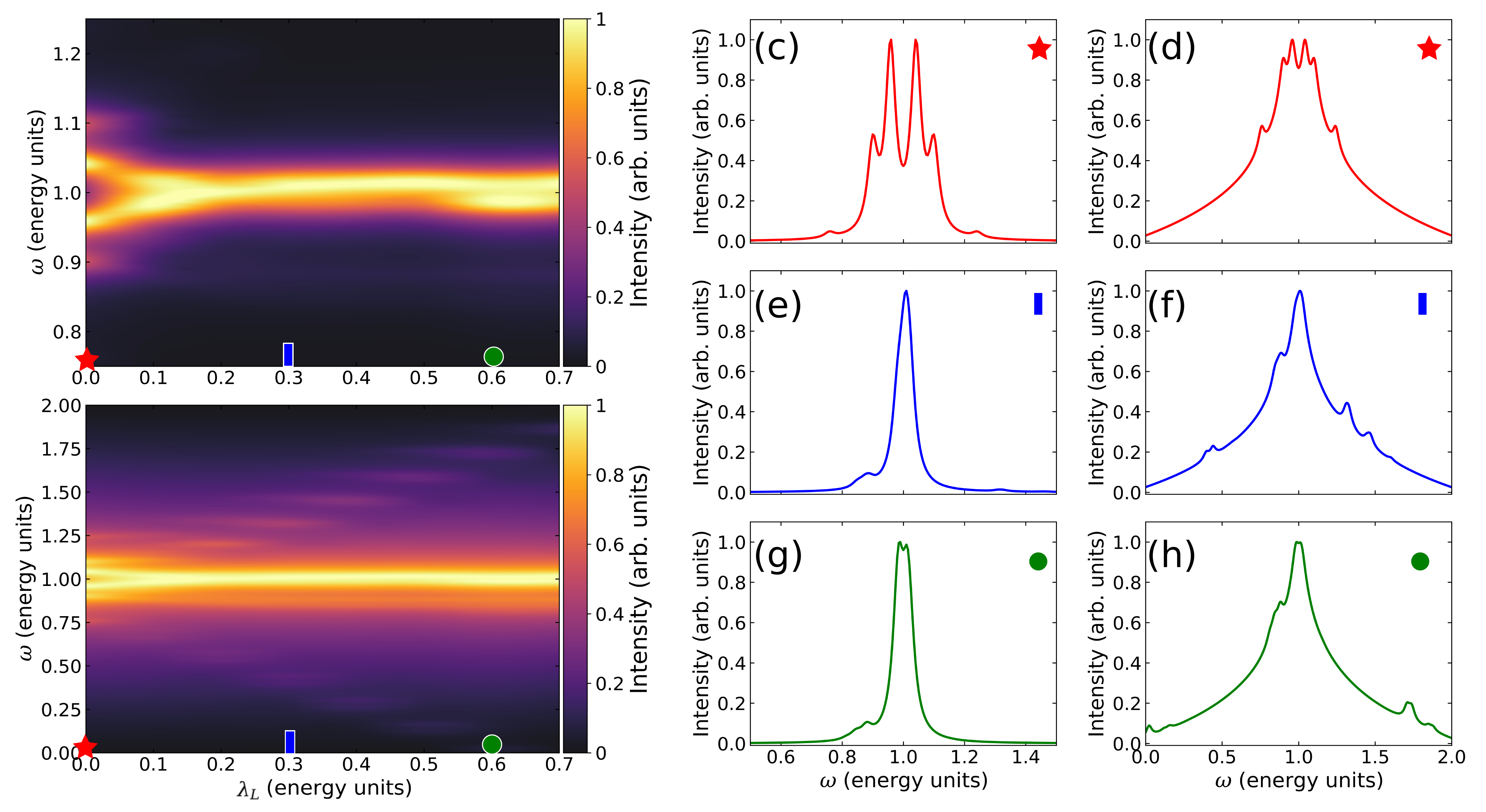}}
	\caption{\label{fig:SpectrumIsolatedMBSs} Normalized emission spectrum [Eq.~(\ref{eq:Spectrum})] for the situation of isolated MBSs localized at opposite nanowire ends ($\lambda_R = \varepsilon_M = 0$), with $\omega_c = \omega_{eg}=1.0$, $\Omega_{R}=0.1$, $P=0.015$ and $\gamma_{\text{ph}}=0.02$. (a) and (b): Intensity of emission spectrum as a function of both emitted photon frequency $\omega$ and QD-left MBS coupling $\lambda_L$, in linear and logarithmic scales, respectively. Panels (c,e,g) are linecuts indicated by geometric shape markers in (a), while panels (d,f,h) correspond to same linecuts, but in logarithmic scale.}
\end{figure*}

\begin{figure*}[t]
	\centerline{\includegraphics[width=6.8in,keepaspectratio]{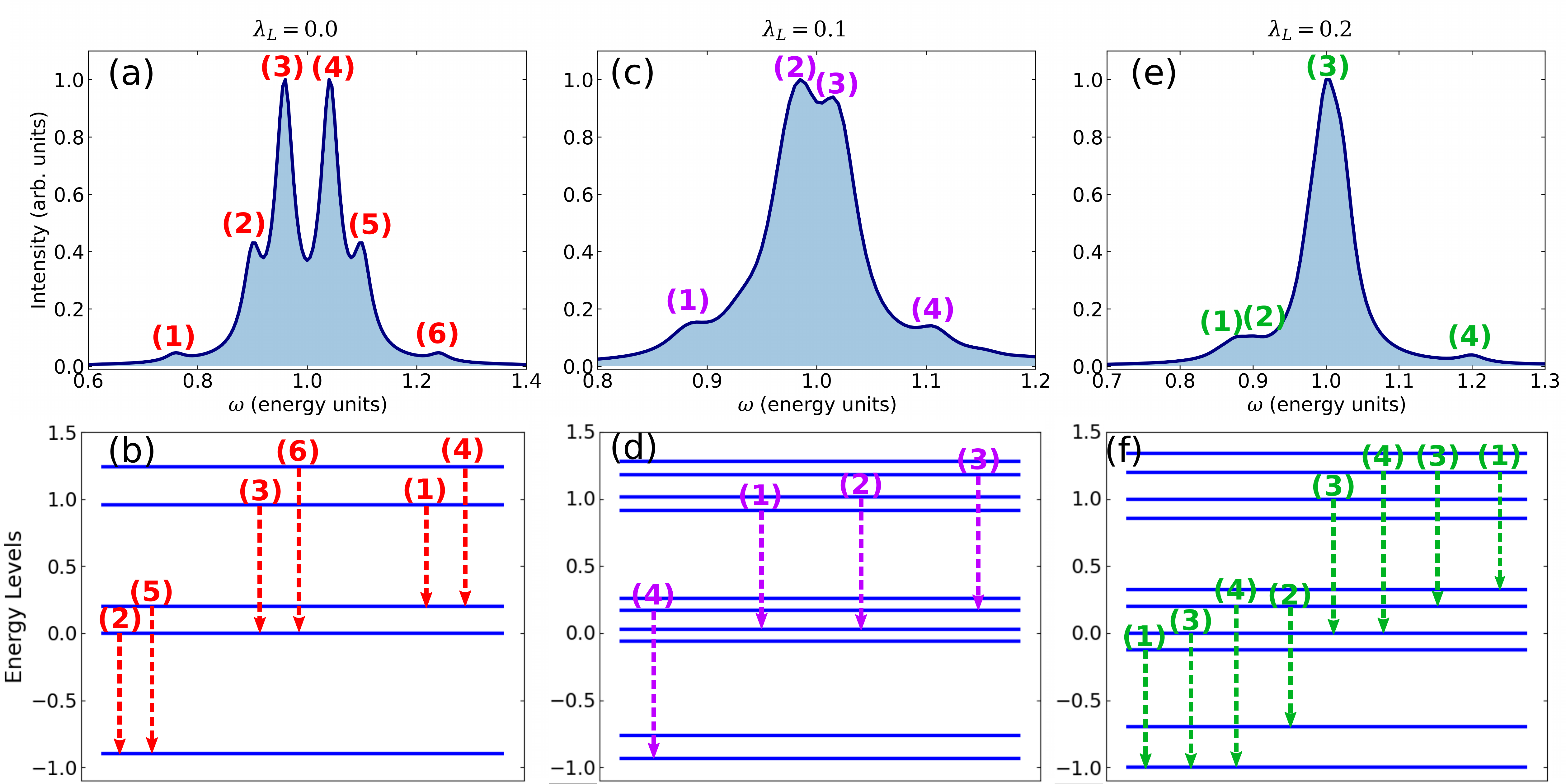}}
	\caption{\label{fig:analysisSpectrumTransitions} Upper panels: normalized emission spectrum [Eq.~(\ref{eq:Spectrum})] as a function of the frequency of the emitted photon $\omega$, for the situation in which the MBSs are well-localized at the nanowire ends and do not overlap with each other ($\lambda_R =\varepsilon_{M} =0.0$), with $\omega_c = \omega_{eg}=1.0$, $\Omega_{R}=0.1$, $P=0.015$ and $\gamma_{\text{ph}}=0.02$, considering distinct values of the QD-left MBS coupling $\lambda_{L}$. Lower panels: allowed optical transitions of the system which correspond to the peaks in the upper panels, labeled by colored numbers.}
\end{figure*}

\begin{figure}[t]
	\centerline{\includegraphics[width=3.6in,keepaspectratio]{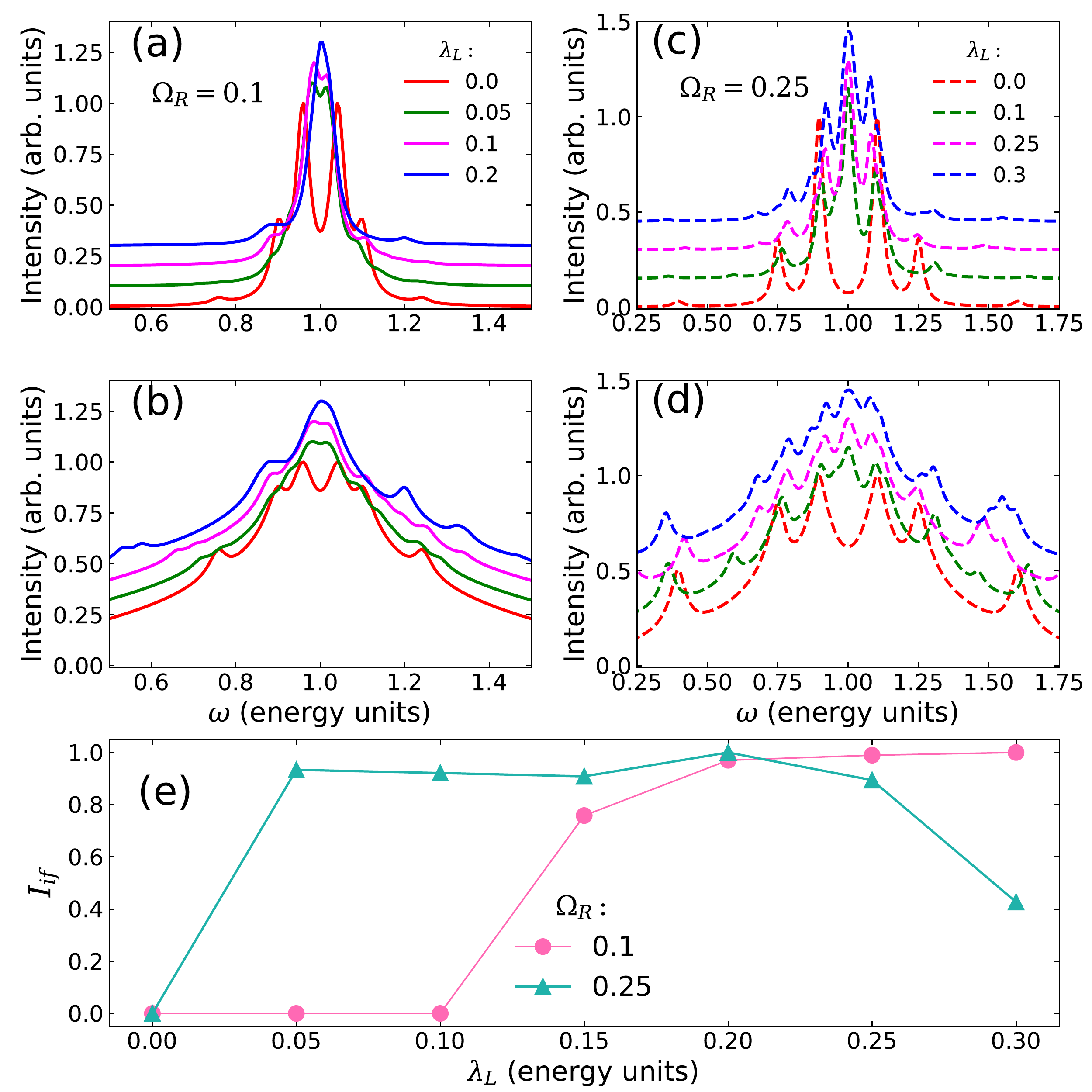}}
	\caption{\label{fig:SpectrumDistictOmegaR}Upper panels: normalized emission spectrum [Eq.~(\ref{eq:Spectrum})] as a function of the emitted photon frequency $\omega$, for the situation where the MBSs are well-localized at the nanowire ends and do not overlap with each other ($\lambda_R =\varepsilon_{M} =0.0$), considering two values of Rabi frequency $\Omega_{R}$, with $\omega_c = \omega_{eg}=1.0$, $P=0.015$, $\gamma_{\text{ph}}=0.02$ and distinct values of QD-left MBS coupling $\lambda_{L}$. Middle panels: Same emission spectrum of upper panels, but in the logarithmic scale. In both upper and middle panels, the plots are slightly shifted in the \textit{y}-axis for a better visualisation. Lower panel: evolution of the normalized transition probabilities [Eq.~(\ref{eq:transitions})] at frequency $\omega \approx \omega_{c}$ as a function of the QD-left MBS coupling $\lambda_{L}$.}
\end{figure}

\begin{figure*}[t]
	\centerline{\includegraphics[width=7.2in,keepaspectratio]{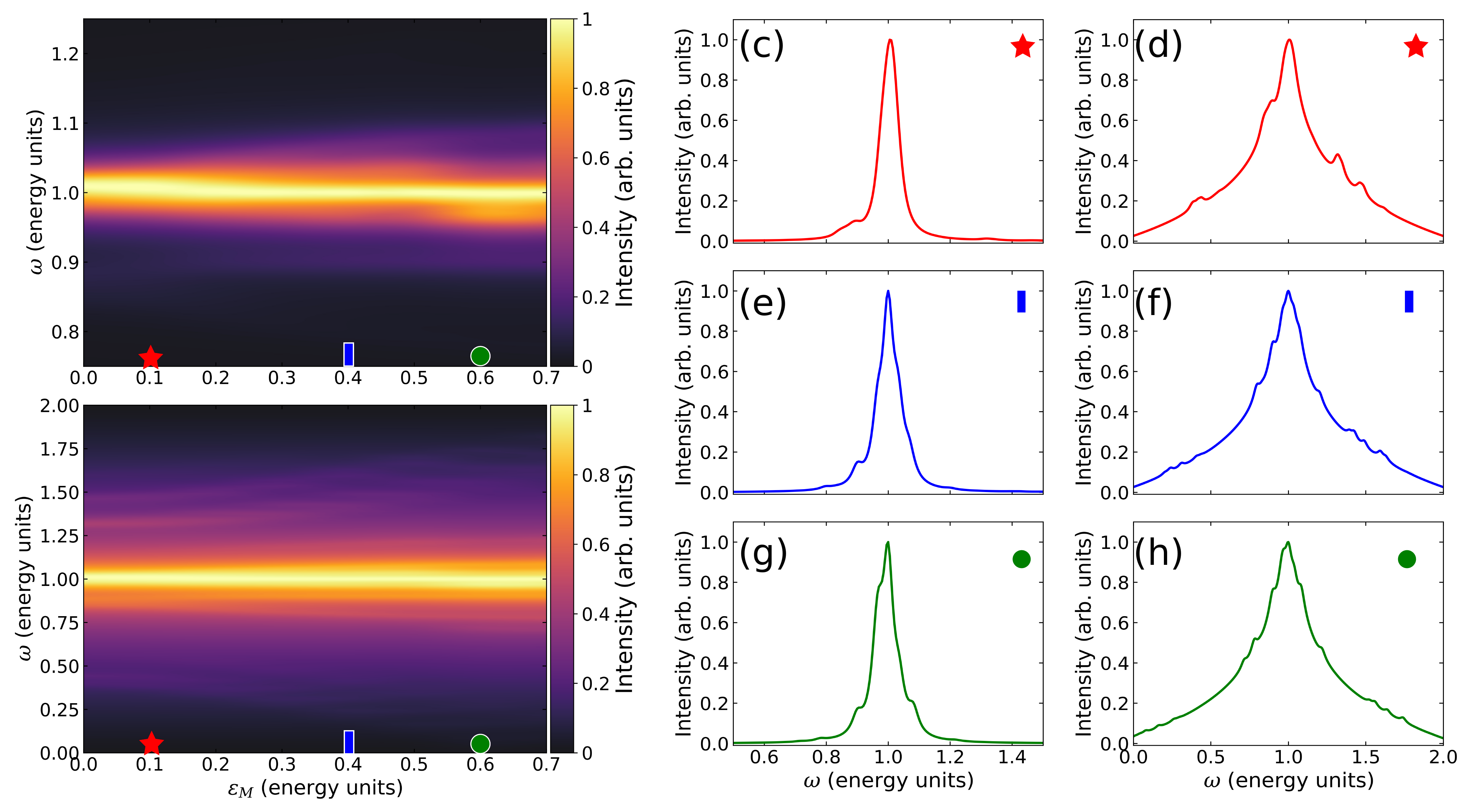}}
	\caption{\label{fig:SpectrumOverlappedMBSs} Normalized emission spectrum [Eq.~(\ref{eq:Spectrum})] for the situation of MBSs localized at opposite nanowire ends ($\lambda_R = 0.0$), but with a finite overlap $\varepsilon_M$ between each other, with $\lambda_L = 0.3$, $\omega_c = \omega_{eg}=1.0$, $\Omega_{R}=0.1$, $P=0.015$ and $\gamma_{\text{ph}}=0.02$. (a) and (b): Intensity of emission spectrum as a function of both emitted photon frequency $\omega$ and left-right MBS overlap $\varepsilon_{M}$, in linear and logarithmic scales, respectively. Panels (c,e,g) are linecuts indicated by geometric shape markers in (a), while panels (d,f,h) correspond to same linecuts, but in logarithmic scale.}
\end{figure*}

\begin{figure*}[t]
	\centerline{\includegraphics[width=7.2in,keepaspectratio]{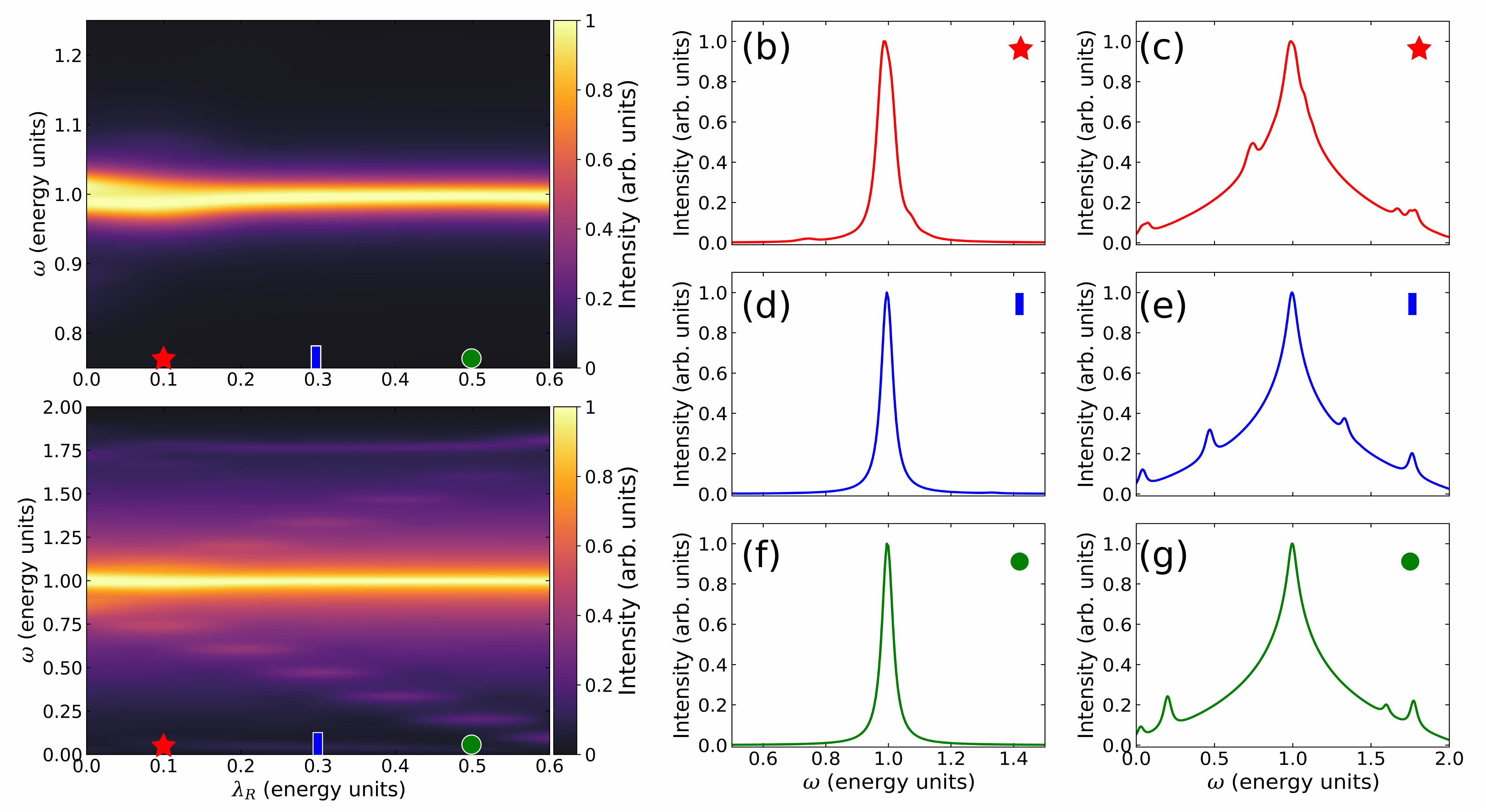}}
	\caption{\label{fig:SpectrumlocalMBSs} Normalized emission spectrum [Eq.~(\ref{eq:Spectrum})] for the situation where the MBSs are not well localized at the nanowire ends ($\lambda_R \neq 0.0$) and have a negligible overlap  between each other ($\varepsilon_{M}\ll \lambda_R,\lambda_L$), with $\lambda_L = 0.6$, $\omega_c = \omega_{eg}=1.0$, $\Omega_{R}=0.1$, $P=0.015$ and $\gamma_{\text{ph}}=0.02$. (a) and (b): Intensity of emission spectrum as a function of both emitted photon frequency $\omega$ and QD-right MBS coupling $\lambda_{R}$, in linear and logarithmic scales, respectively. Panels (c,e,g) are linecuts indicated by geometric shape markers in (a), while panels (d,f,h) correspond to same linecuts, but in logarithmic scale.}
\end{figure*}

The standard way of computing both the external incoherent pumping $P$ and decay of photons $\gamma_{\text{ph}}$ from the closed system to the reservoir is obtaining the full density matrix $\rho$ of the system through the master equation~\cite{MicrocavitiesBook}
\begin{equation}
\partial_{t}\rho = -\frac{\imath}{\hbar}[\hat{\mathcal{H}},\rho] + \mathcal{L}\rho,\label{eq:MasterEquation}    
\end{equation}
where $\mathcal{L}$ is the Lindblad superoperator. By solving numerically Eq.~(\ref{eq:MasterEquation}), the density matrix in the steady-state (SS) regime $\rho^{\text{SS}}$ can be found, which provides the information about the occupations of different quantum states of the system for a given $P$ and $\gamma_{\text{ph}}$. 

The emission spectrum in the SS regime can be numerically obtained by a modified Fermi golden rule~\cite{Savenko2012} and reads

\begin{equation}
S(\omega)\sim \varrho(\omega) \sum_{i,f}\rho_{ii}^{\text{SS}}I_{if}\frac{\gamma_{\text{ph}}^{2}}{(\varepsilon_{i}-\varepsilon_{f} - \omega)^2+\gamma_{\text{ph}}^{2}}.\label{eq:Spectrum}
\end{equation}
wherein $\varrho(\omega)\propto \omega^{3}$ is the reservoir density of states, $I_{if}$ [Eq.~(\ref{eq:transitions})] are the transition probabilities between the initial and final states of the system Hamiltonian associated with photon emission to the reservoir, with the corresponding eigenenergies $\varepsilon_{i,f}$, and $\rho_{ii}^{\text{SS}}$ are probabilities to find the system in the eigenstates of $\hat{\mathcal{H}}$ which should be obtained by solving master equation [Eq.~(\ref{eq:MasterEquation})].

\section{Results and Discussion}\label{Results and Discussion}

Below, we numerically compute the optical transitions probabilities [Eq.(\ref{eq:transitions})] for a QD embedded in the microcavity and coupled to the Majorana nanowire ($\lambda_{L}\neq 0$), considering the cases (i), (ii) and (iii) previously discussed in Sec.~\ref{Hamiltonian}. We set the following parameters for the QD embedded in the microcavity: energy of QD excited state $\omega_{e}=0.1$, energy of QD ground state $\omega_{g}=-0.9$, Rabi frequency $\Omega_R = 0.1$ and eigenfrequency of the cavity photons $\omega_c = \omega_{e} -\omega_{g} = 1.0$, all in energy arbitrary units. The numerical calculations were carried out using \textit{QuTiP 4.6.2 package (Python 3.8.10)}~\cite{qutip1,qutip2}. Moreover, in all the results we use the distance between ground and excited states as an energy unit.

Fig.~\ref{fig:Transitions} shows the allowed optical transitions, as well as the evolution of the first $20$ eigenenergies of the Hamiltonian [Eq.~(\ref{eq:Hfull})] for several values related to the QD-Majorana nanowire parameters $\lambda_{L}$, $\lambda_{R}$ and $\varepsilon_{M}$. The allowed transitions between the eigenenergies are represented by red arrows. Fig.~\ref{fig:Transitions}(a) specifically exhibits the behavior of the energy spectrum of the composite system (cavity photons-QD-Majorana nanowire) as the QD-left MBS coupling $\lambda_{L}$ is increased, for the case wherein the MBSs are well localized at the nanowire ends, totally apart from each other (i). It can be easily seen that a finite $\lambda_{L}$ leads to the appearance of new energy levels which are not present in the original JC ladder. Moreover, the energy levels which belongs to the JC model, are renormalized, leading to a shift in the original rungs of the JC ladder. This reshape of energy levels allows new optical transitions if compared to the allowed transitions in the JC model~\cite{Savenko2012,MicrocavitiesBook,Bruce1993}, as shown in Fig.~\ref{fig:Transitions}(b) for $\lambda_{L}=0.3$. It can be noticed that there are transitions between two nearest energy levels and between two more distant levels. These transitions will be responsible for an emitted radiation in regions far from $\omega_c$, as we shall see later on.

Fig.~\ref{fig:Transitions}(c) depicts the evolution of the system energy spectrum as the overlap strength $\varepsilon_M$ between the MBSs bound at opposite ends of the Majorana nanowire is enhanced (ii) (we keep $\lambda_L = 0.3$). It can be noticed that the effect of a finite $\varepsilon_M$ is breaking the degeneracy of certain eigenenergies, which consequently leads to additional rungs in the ladder, if compared to Fig.~\ref{fig:Transitions}(b). The reshape of the energy spectrum due to the overlap between left and right MBSs also allows new optical transitions between nearest and more distant rungs, as shown in Fig.~\ref{fig:Transitions}(d), where $\varepsilon_{M}=0.4$ [light green rectangle in Fig.~\ref{fig:Transitions}(c)].

The case in which the right MBS are not well localized at the right nanowire end (iii), leading to a finite coupling with the QD ($\lambda_{R}\neq 0$), is illustrated in Fig.~\ref{fig:Transitions}(e). The corresponding energy spectrum evolution for increasing values of $\lambda_{R}$, with $\lambda_{L}=0.6$ and $\varepsilon_{M}=10^{-3}$, reveals that the finite QD-right MBS hybridization also leads to a degeneracy breaking if compared to the case of isolated MBSs [Fig.~\ref{fig:Transitions}(a)]. As in the previous case of overlapped MBSs [Fig.~\ref{fig:Transitions}(d)], it is observed in Fig.~\ref{fig:Transitions}(f) the emergence of new rungs in the ladder, also allowing new optical transitions. But distinct from the overlapped case, the coupling between the QD and the right MBS opens transitions between extremely close ladder rungs, which will be responsible for near zero-frequency peaks in the emission spectrum profile for any value of $\lambda_{R}$, as we shall see further on. 

In the following figures, the emission spectrum [Eq.~(\ref{eq:Spectrum})] behavior is analyzed in detail for the same set of QD-microcavity parameters adopted in Fig.~\ref{fig:Transitions}, considering the incoherent pumping intensity $P=0.015$ and the cavity photons decay rate $\gamma_{ph}=0.02$. For simplicity, we take the excitonic decay rate $\gamma_{QD}$ to be zero. Notice that the regime of strong light-matter coupling $\Omega_R > |\gamma_{ph}-\gamma_{QD}|/4$~\cite{MicrocavitiesBook} is fulfilled.

The emission spectrum for the situation of isolated and well localized MBSs at the nanowire ends (i) is shown in Fig.~\ref{fig:SpectrumIsolatedMBSs}. Panel (a) depicts the normalized emission spectrum as a function of both the QD-left MBS coupling $\lambda_L$ and emitted photon frequency $\omega$. For $\lambda_L = 0$ (red star marker), the emission spectrum shows a quadruplet pattern around the cavity photons eigenfrequency $\omega = \omega_c$, as shown in Fig.~\ref{fig:SpectrumIsolatedMBSs}(c). This quadruplet structure is expected for the JC model in the strong-coupling regime in presence of a small incoherent pump~\cite{Savenko2012,delVallePhysRevB79235326(2009)}. This multiplet structure, also known as JC fork, are also shown on a logarithmic scale, as depicted in Figs.~\ref{fig:SpectrumIsolatedMBSs}(b) and (d). 

Fig.~\ref{fig:SpectrumIsolatedMBSs}(a) also reveals that as $\lambda_L$ is increased, the multiplet structure coalesces into single peak at $\omega=\omega_c$, with a smaller peak structure at $\omega \approx 0.9$. This is better seen in Fig.~\ref{fig:SpectrumIsolatedMBSs}(e), with corresponds to $\lambda_L = 0.4$ [blue rectangle marker in (a)]. For bigger values of QD-left MBS coupling, the single-peak structure at $\omega = \omega_{c}$ undergoes a tiny splitting, but still preserves its single-peak shape, as shown in Fig.~\ref{fig:SpectrumIsolatedMBSs}(g).

In addition to the evolution of the JC fork to a single peak shown in Fig.~\ref{fig:SpectrumIsolatedMBSs}(a), the leakage of the isolated left MBS also drives the formation of weaker peaks in the emission spectrum far from the cavity eigenfrequency region, which are visible only on a logarithmic scale, as depicted in Fig.~\ref{fig:SpectrumIsolatedMBSs}(b) and the corresponding linecuts shown in Figs.~\ref{fig:SpectrumIsolatedMBSs}(f)-(h). Figs.~\ref{fig:SpectrumIsolatedMBSs}(h), for instance, shows a finite emitted radiation at very-low frequency $\omega$ close to the Rabi frequency, as well as some peaks near the double frequency $\omega = 2\omega_c$. The low frequency signal can be amplified by placing the QD-microcavity inside a $\SI{}{\tera\hertz}$ frequency cavity, which will resonantly enhance the density of states $\rho(\omega)$ and increase the emission up to a factor of $10$~\cite{TodorovPRL99223603(2007),SIZOV2010THzDetectors}.

In Fig.~\ref{fig:analysisSpectrumTransitions}, we investigate in detail the evolution of the normalized emission spectrum as a function of the emitted photon frequency $\omega$, visible on linear scale, and the corresponding optical transitions responsible for the emission spectrum peaks as $\lambda_{L}$ is turned on, considering the same parameters adopted in Fig.~\ref{fig:SpectrumIsolatedMBSs}. Panel (a) shows the emission spectrum when the QD-microcavity is decoupled from the Majorana nanowire ($\lambda_{L}=0$), wherein the well-known JC fork structure appears, with a well-resolved double-peak around the cavity eigenfrequency $\omega_{c}=1.0$. This multiplet profile comes from the JC ladder shown in panel (b)~\cite{Savenko2012,delVallePhysRevB79235326(2009)}, in which each optical transition between the system eigenenergies corresponds to a peak in (a), as labeled by the red numbers.

Fig.~\ref{fig:analysisSpectrumTransitions}(c) depicts the emission spectrum for $\lambda_{L}=0.1$. We can notice that the JC fork lineshape from Fig.~\ref{fig:analysisSpectrumTransitions}(a) is deformed and the well-resolved peaks around $\omega=\omega_{c}$ becomes merged into each other. This changing in the emission spectrum profile comes from the emergence of new system eigenenergies, or equivalently, new rungs in the JC ladder, due to the leaking of the isolated left MBS into the QD, which opens new optical transitions, as depicted in Fig~\ref{fig:analysisSpectrumTransitions}(d). The peaks in the emission spectrum around $\omega_c$ in panel (c), for instance, comes from the transitions (2) and (3) shown in panel (d)

For $\lambda_{L}=0.2$, Fig.~\ref{fig:analysisSpectrumTransitions}(e) shows that the JC fork structure is completely absent, with the appearance of a single peak at $\omega = \omega_{c}=1.0$. This single peak indicates that optical transitions with the frequency on resonance with the cavity eigenfrequency are now allowed in the system. Particularly, these transitions are labeled by the green number (3) in Fig.~\ref{fig:analysisSpectrumTransitions}(f) and the other optical transitions yields the satellite peaks in panel (e). In addition, comparison between Figs.~\ref{fig:analysisSpectrumTransitions}(d) and (f) suggests that the increasing of $\lambda_{L}$ moves each closest pair of ladder rungs (system eigenenergies) apart from each other [see also Fig.~\ref{fig:Transitions}], thus opening the possibility of new optical transitions in distinct frequencies and hence, changing the corresponding emission spectrum profile as a function of $\omega$.

In Fig.~\ref{fig:SpectrumDistictOmegaR}, we particularly analyze the competition of $\lambda_{L}$ and $\Omega_{R}$ energy scales in the mutation of the doublet structure, formed very near to $\omega = \omega_{c}$ in the emission spectrum for $\lambda_{L}=0$ [Fig.~\ref{fig:analysisSpectrumTransitions}(a)], to a single peak localized at $\omega=\omega_{c}$ when $\lambda_{L}\neq 0$ [Fig.~\ref{fig:SpectrumIsolatedMBSs}(e) and Fig.~\ref{fig:analysisSpectrumTransitions}(e)], considering the same other parameters adopted in Fig.~\ref{fig:SpectrumIsolatedMBSs}.

Figs.~\ref{fig:SpectrumDistictOmegaR}(a) and (b) exhibit the emission spectrum as a function of the emitted photon frequency for distinct values of $\lambda_{L}$, in linear and logarithmic scales, respectively, considering the same value of Rabi frequency $\Omega_{R}=0.1$ adopted throughout this work. Both the panels indicate that the well resolved doublet structure around $\omega=\omega_{c}$ for $\lambda_{L}=0$ (red line) transforms into an almost single-peak, with two peaks merging with each other for $\lambda_{L}\leq\Omega_{R}$ (green and magenta lines). For the biggest value of $\lambda_{L}$ adopted in Fig.~\ref{fig:SpectrumDistictOmegaR}(a)-(b) (blue line), a well-defined single peak structure is seen at $\omega=\omega_{c}$ due to new allowed optical transitions in the system, as discussed earlier. The positions of the satellite peaks also changes for distinct values of $\lambda_{L}$, as better visualized in Fig.~\ref{fig:SpectrumDistictOmegaR}(b).

To investigate if the well-resolved doublet structure around $\omega=\omega_{c}$ is indeed modified for $\lambda_{L}\leq\Omega_{R}$, in Figs.~\ref{fig:SpectrumDistictOmegaR}(c) and (d) we consider $\Omega_{R}=0.25$. For this value of Rabi frequency, a single peak already arises at $\omega_c$ for $\lambda_{L}<\Omega_{R}$ (dashed green line) and remains for bigger values of $\lambda_{L}$ adopted. The emission spectrum in the logarithmic scale shows a shift in the positions of satellite peaks with $\lambda_{L}$, as also observed in Fig.~\ref{fig:SpectrumDistictOmegaR}(b).

The findings shown in Figs.~\ref{fig:SpectrumDistictOmegaR}(a)-(d) can be explained by Fig.~\ref{fig:SpectrumDistictOmegaR}(e), which depicts the behavior of the normalized transition probabilities of the system [Eq.~(\ref{eq:transitions})] at the cavity eigenfrequency $\omega\approx\omega_c$ as a function of $\lambda_{L}$, for both the values of $\Omega_R$ adopted in upper panels. For $\Omega_{R}=0.1$, there is no optical transition allowed at the cavity eigenfrequency when $\lambda_{L}\leq \Omega_{R}$. This is the reason why we see merging peaks very near to $\omega=\omega_c$ instead of a well defined single peak in Fig.~\ref{fig:SpectrumDistictOmegaR}(a)-(b) when $\lambda_{L}=0.05$ and $\lambda_{L}=0.1$, although the characteristic well-resolved doublet structure around the cavity eigenfrequency for $\lambda_{L}=0$ is suppressed. When the Rabi frequency is increased for $\Omega_{R}=0.25$, Fig.~\ref{fig:SpectrumDistictOmegaR}(e) reveals that $\lambda_{L}\leq \Omega_{R}$ is enough for allowing optical transitions with frequency on resonance with the cavity eigenfrequency, thus yielding well-defined peaks at $\omega=\omega_{c}$ in the emission spectrum. Nevertheless, Fig.~\ref{fig:SpectrumDistictOmegaR} shows that a small leaking strength of the left MBS into the QD $\lambda_{L}\leq \Omega_{R}$ spoils the well resolved emission spectrum doublet structure around $\omega=\omega_{c}$, leading to either merging peaks or a well-defined single peak at the cavity eigenfrequency, being this last one due to finite transition probabilities at this frequency.

Fig.~\ref{fig:SpectrumOverlappedMBSs}(a) exhibits the normalized emission spectrum as a function of the emitted photon frequency $\omega$ and the overlap strength $\varepsilon_{M}$ between the MBSs at opposite ends of the nanowire (ii), for $\lambda_{L} = 0.3$. It is noticed that the increasing of $\varepsilon_{M}$ does not change the the single-peak structure at $\omega_{c}=1.0$ present in the previous case of isolated MBSs (Fig.~\ref{fig:SpectrumIsolatedMBSs}). The persistent central peak in presence of a finite MBS-MBS overlap is also shown in the line cuts of Figs.~\ref{fig:SpectrumOverlappedMBSs}(c), (e) and (g), respectively.

The difference between the current case of finite $\varepsilon_M$ and the previous case of isolated MBSs is unveiled in the logarithmic scale, as exhibited in Fig.~\ref{fig:SpectrumOverlappedMBSs}(b) and the corresponding linecuts in (d), (f) and (h). It is seen the emergence of several peaks near to each other with very low amplitude, even on a logarithmic scale. These low-amplitude multi-peaks arise owing to the opening of new optical transitions caused by the the degeneracy breaking of the energy levels due to $\varepsilon_{M}\neq 0$ [Fig.~\ref{fig:Transitions}(c,d)].

The situation (iii) wherein the right MBS are not well localized at the nanowire end ($\lambda_{R} \gg \varepsilon_{M}$) is illustrated in Fig.~\ref{fig:SpectrumlocalMBSs}, for $\lambda_L = 0.6$. Panel (a) reveals the emission spectrum behavior as a function of both the emitted photon frequency $\omega$ and QD-right MBS coupling $\lambda_{R}$. As in the previous case of overlapped MBSs localized at the nanowire ends (Fig.~\ref{fig:SpectrumOverlappedMBSs}), the single peak structure at the cavity eigenfrequency $\omega_{c}=1.0$ still remains, even for higher values of $\lambda_{R}$. The single-peak in the emission spectrum also can be seen in the corresponding linecuts shown in Figs.~\ref{fig:SpectrumlocalMBSs}(c), (e) and (g). 

The difference between the current case and the previous situations of highly isolated MBSs [Fig.~\ref{fig:SpectrumIsolatedMBSs}(b)] and overlapped MBSs [Fig.~\ref{fig:SpectrumOverlappedMBSs}(b)] is observed in the emitted radiation of low and high frequency, and thus it is only visible on a logarithmic scale, as shown in Fig.~\ref{fig:SpectrumlocalMBSs}(b). The most striking difference is the presence of a peak at very-low frequency ($\omega \approx 0.1$), as well as a peak near the region of double-frequency, precisely at $\omega \approx 1.75$, which remain for all values of $\lambda_R$ adopted, as depicted in the corresponding linecuts shown in Figs.~\ref{fig:SpectrumlocalMBSs}(d), (f) and (h), respectively. The emitted radiation at very-low frequency comes from allowed optical transitions between the nearest energy levels of the corresponding ladder shown in Fig.~\ref{fig:Transitions}(f), for instance. Oppositely, the emitted radiation in the region around the $\omega \approx 1.75$ appears due to transitions between the farthest rungs of the corresponding ladder.      

\section{Conclusions}\label{Conclusions}

We have analyzed the optical response of a quantum dot (QD) embedded in a microcavity and coupled to a Majorana nanowire hosting Majorana bound states (MBSs) at its opposite ends. In the regime of the strong light matter coupling, we demonstrated that the coupling between the Majorana nanowire and the QD opens new optical transitions between the polaritonic states of the QD-microcavity system, thus reshaping the well-known JC ladder. Moreover, we also demonstrated that the reshaping of the ladder rungs and corresponding optical transitions depends on the spatial location of the MBSs with respect to the nanowire ends, as well as the overlap between them. Consequently, the opening of new transitions strongly affects the cavity emission spectrum, which shows an asymmetric pattern having a prominent single-peak structure centered at the cavity eigenfrequency for all the cases explored, namely, highly isolated MBSs localized at the nanowire ends, overlapped MBSs and the right MBS displaced from its corresponding nanowire end. The distinction between the three situations is observed only in the emitted radiation of low and high frequency.   
\begin{acknowledgments}
This work was supported by the Icelandic Research Fund (project ``Hybrid polaritonics''). VKK acknowledges the support from the Georg H. Endress foundation. ACS acknowledges support from Brazilian National Council for Scientific and Technological Development (CNPq), grants~305668/2018-8 and 308695/2021-6. IAS acknowledges support from the Ministry of Science and Higher Education of Russian Federation, goszadanie no. 2019-1246 and Priority 2030 Federal Academic Leadership Program. 
 
\end{acknowledgments}




\bibliography{apssamp}

\end{document}